\documentclass[aps,twocolumn,showpacs,prl,reprint]{revtex4}
\usepackage{graphicx}
\usepackage{color}

\newcommand{\ket}[1]{|#1\rangle}

\begin{document}

\title{Comparison of Atom Interferometers and Light Interferometers as Space-Based Gravitational Wave Detectors}
\author{John G. Baker and J.I. Thorpe}
\affiliation{Gravitational Astrophysics Laboratory, NASA Goddard Space Flight Center, 8800 Greenbelt Rd., Greenbelt, MD 20771, USA}

\begin{abstract}
We consider a class of proposed gravitational wave detectors based on multiple atomic interferometers separated by large baselines and referenced by common laser systems. We compute the sensitivity limits of these detectors due to intrinsic phase noise of the light sources, non-inertial motion of the light sources, and atomic shot noise and compare them to sensitivity limits for traditional light interferometers. We find that atom interferometers and light interferometers are limited in a nearly identical way by intrinsic phase noise and that both require similar mitigation strategies (e.g. multiple arm instruments) to reach interesting sensitivities. The sensitivity limit from motion of the light sources is slightly different and favors the atom interferometers in the low-frequency limit, although the limit in both cases is severe. 

\end{abstract}
\maketitle

\emph{Introduction.---}The detection and measurement of gravitational waves (GWs) from astrophysical and cosmological sources is recognized as one of the most promising sources for new information about the universe\cite{Schutz_99} and has been a goal of experimental physicists for nearly half a century\cite{Weber60,Weber67,Hellings78,Detweiler79,Abromovici92,Tinto98b,Hobbs10}. The milliHertz-frequency region of the GW spectrum is expected to be particularly rich in GW sources and has been the target of proposed space-based instruments, most notably the Laser Interferometer Space Antenna (LISA) \cite{Bender_98}. LISA was identified as a priority in the most recent decadal survey of astronomy and astrophysics\cite{Astro2010} but has yet to be implemented due to funding constraints.

GW detectors based on a single isolated atom interferometer (AI) have been considered\cite{Chiao04} but found to have little advantage over light interferometers\cite{Roura06, Delva06}. More recent proposals use two AIs separated by a large baseline and referenced to a common pair of lasers\cite{Dimopoulos08b,Hogan2011}. These instruments use the AIs both to provide an inertial reference and to measure the phase of the light fields used as the atom ``optics''. When the light source is placed sufficiently far from the AI, the optical phase contains a non-negligible contribution from GWs. However, the optical phase measured by the AI also contains contributions from intrinsic phase fluctuations of the light source and Doppler motion of the light source relative to the atoms. By using a common pair of lasers for both AIs, the proposed GW detectors can eliminate contributions from \emph{one} of these light sources. However, the contributions from the second light source remain. In this letter, we calculate the limits on GW sensitivity resulting from intrinsic phase fluctuations and light source motion for a GW detector consisting of two AIs while making a parallel analysis of a light interferometer (analogous to a single `arm' of LISA). 

Our treatment clarifies the relative merits of the two approaches to space-based GW measurement and may be helpful in future GW instrument design considerations.

\begin{figure}[!ht]
\includegraphics[height=2.5in]{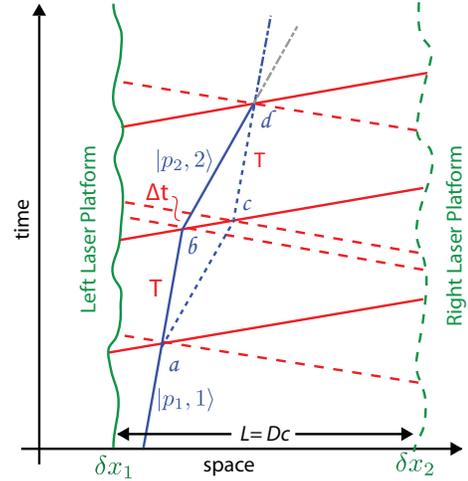}
\caption{\label{fig:oneMZ}A three-pulse Mach-Zehnder atom interferometer controlled by two lasers on separate platforms separated by a distance $L$}
\end{figure}

\emph{Analysis.---}
A single three-pulse Mach-Zehnder (MZ) AI, like the one shown in Figure \ref{fig:oneMZ}, is controlled by two lasers (``left'' and ``right'') on separate platforms separated by a distance $L$. A three-level atomic system is assumed with ground states $\ket{p_i,i}$, where $p_i$ describes the linear momentum of the atom in the $\hat{x}$-direction and $i=1,2$ denotes the internal state.  An atom cloud prepared in state $\ket{p_1,1}$ enters the interferometer and is subjected to a Raman $\frac{\pi}{2}$-pulse beam splitter at point \emph{a} at time $t-2T$. This splits the atom wavepacket into a portion in state $\ket{p_1,1}$ and a portion in state $\ket{p_2,2}$. At a time $t-T$, a $\pi$-pulse converts the $\ket{p_1,1}$ portion into $\ket{p_2,2}$ at point \emph{b}. A short time $\Delta t$ later, the same pulse converts the $\ket{p_2,2}$ state back into a $\ket{p_1,1}$ state at point \emph{c}. At time $t$, the two wave function paths converge (point \emph{d}) and are re-combined with another $\frac{\pi}{2}$ pulse.

After recombination, the population in one or both of the two ground states is measured and the result is used to determine the phase of the wavefunction. It has been shown\cite{Kasevich92} that the wavefunction phase response, $\gamma(t)$, from the AI measured just after the third pulse at time $t$ is given by
\begin{equation}
\label{eq:gamma}
\gamma(t)\approx\delta\Phi(t)-2\delta\Phi(t-T)+\delta\Phi(t-2T)+\delta\Phi_{m}(t),
\end{equation}
where $\delta\Phi(t)$ is 
the difference from nominal phases of the optical phases observed by the atoms and $\delta\Phi_{m}(t)$ is any measurement noise in the AI, for instance atom shot noise. We assume $T\gg \Delta t$ and we neglect the finite duration of the Raman pulses here for clarity \cite{Cheinet08}. In the Fourier domain, (\ref{eq:gamma}) becomes
\begin{equation}
\label{eq:gammaf}
\tilde{\gamma}=-4 \sin^2\left(\omega T/2\right) e^{-i\omega T}\delta\tilde{\Phi}+\delta\tilde{\Phi}_{m},
\end{equation}
where tilde denotes the Fourier transform of a timeseries and $\omega$ is the angular Fourier frequency. In the low frequency limit $\omega T \ll 1$ the transfer function from $\delta\tilde{\Phi}$ to $\tilde{\gamma}$ is that of a second time derivative. The significant time-varying contributions to the 
phase differences are given by
\begin{equation}
\delta\Phi(t)=\delta\phi_L(t)-\delta\phi_R(t),
\end{equation}
where $\delta\phi_L(t)$ are the phase variations of the left light source retarded from a reference point in the vicinity of the atom cloud, and $\delta\phi_R(t)$ are the retarded phase variations of the right light source. Specifically,
\begin{eqnarray}
\delta\phi_{L}(t)&\approx&k\delta x_1(t)+\delta\phi_1(t),\\
\delta\phi_{R}(t)&\approx&-k\delta x_2(t-D)+\delta\phi_2(t-D)+kc Y_-(t),
\end{eqnarray}
where $k$ is the nominal wavenumber of the photons, neglecting the splitting between the $\ket{1}$ and $\ket{2}$ internal states and other higher-order effects. The approximate light
propagation time between the distant right laser and the atom cloud is $D=L/c$, the position
fluctuations caused by non-gravitational forces on the left and right
laser platforms are given by $\delta x_1(t)$ and $\delta x_2(t)$, and the
intrinsic optical phase noise for the left and
right light sources are given by $\delta\phi_1(t)$ and
$\delta\phi_2(t)$. GWs will cause the received optical phase to differ from the emitted phase by a Doppler shift $Y_-(t)$. 

We ignore any GW effect on the individual AI, assuming the atom separation is much smaller than $L$\cite{Dimopoulos08b}. Similarly, we neglect GW effects on the phase of the left light source, which is assumed to be close to the AI. In the Fourier domain, the phase difference needed for Eq. (\ref{eq:gamma}) is

\begin{equation}
\label{eq:PhiL}
\delta\tilde\Phi_L\approx\left[k\delta\tilde{x}_2-\delta\tilde{\phi}_2\right] e^{-i\omega D}+k\delta\tilde{x}_1+\delta\tilde{\phi}_1-kc \tilde{Y}_-.
\end{equation}

For comparison, we can evaluate the phase difference measured by a one-way light interferometer link, which form the basis of light interferometer detectors such as LISA:

\begin{equation}
\label{eq:PhiL_L}
\delta\tilde\Phi_L^{(l)}\approx\left[k\delta\tilde{x}_2-\delta\tilde{\phi}_2\right] e^{-i\omega D} -k\delta\tilde{x}_1+\delta\tilde{\phi}_1-kc \tilde{Y}_-.
\end{equation}

The primary difference between the phase measured by the AI and that measured by a light interferometer is in the sensitivity to motions of the light sources. In the case of the AI, the measured phase is sensitive to the common-mode motion of the light sources, where in the case of the light interferometer the measured phase is sensitive to differential motion. Importantly, the sensitivity to intrinsic phase noise and gravitational waves are identical.

A classic application of a single AI like that in Figure \ref{fig:oneMZ} is as a gravimeter or accelerometer.
In that case, with the appropriate approximations
$\delta\tilde{x}_1=\delta\tilde{x}_2=\delta\tilde{x}$ (common rigid optics platform), $\delta\tilde{\phi}_1=\delta\tilde{\phi}_2$ (common laser source), $D\approx0$ (short distance), and $\tilde{Y}_-\approx 0 $
, Eq.~\ref{eq:PhiL} reduces to $\delta\tilde{\Phi}_L\approx 2k\delta\tilde{x}$.  Applying the low-frequency limit of (\ref{eq:gammaf}), the output of the MZ is $\tilde{\gamma}\approx -2kT^2\delta\tilde{a}$, where $\delta\tilde{a}\equiv\omega^2\delta\tilde{x}$ is the acceleration noise of the light source (the atoms are assumed to be in an inertial frame in this analysis). This is consistent with results in the literature for AI-based gravimeters\cite{Kasevich91}.

\begin{figure}[ht]
\includegraphics[width=3.25in]{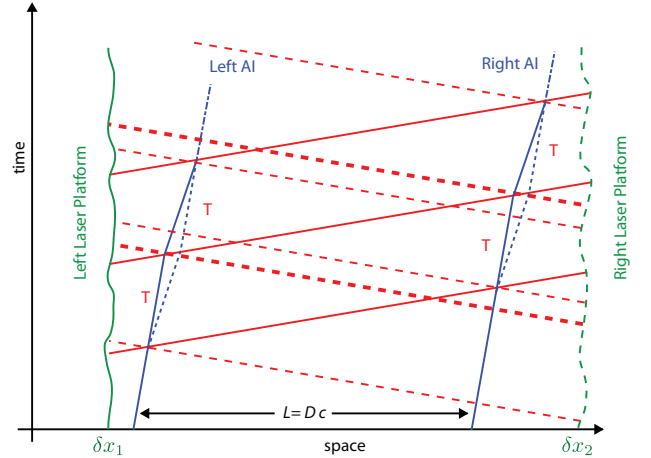}
\caption{\label{fig:AIGW}An arrangement of two three-pulse Mach-Zehnder atom interferometers separated by a baseline L and using common light sources have been proposed as a detector of gravitational waves}
\end{figure}

We consider one-arm GW detectors\cite{Dimopoulos08b, Hogan2011} based on a pair of AIs in an arrangement similar to the design of gravity gradient experiments\cite{Snadden98}, as shown in Figure \ref{fig:AIGW}. A common pair of lasers drives two three-pulse MZ AIs are spaced by a distance $L=cD$.

The response of the left-hand interferometer is given by (\ref{eq:PhiL}).  The response of the right-hand interferometer is similar,

\begin{equation}
\label{eq:PhiR}
\delta\tilde{\Phi}_R\approx\left[k\delta\tilde{x}_1+\delta\tilde{\phi}_1\right] e^{-i\omega D}+k\delta\tilde{x}_2-\delta\tilde{\phi}_2+kc \tilde{Y}_+.
\end{equation}
where $\delta\tilde{\Phi}_R$ denotes the phase difference at an AI vertex on the right atom interferometer and $\tilde{Y}_+$ is the GW effect incurred by the left laser beam propagating to the right atom cloud.

The effect of GWs on electromagnetic propagation from a distant source is familiar from the analysis of space-based light interferometer concepts, pulsar-timing searches for GWs, and microwave Doppler tracking of spacecraft.
The result is a Doppler-shift on the electromagnetic frequency given by \cite{Estabrook75,Hellings81}
\begin{equation}
\label{eq:Dop}
y(t)\equiv\frac{\delta\nu}\nu = \frac12\frac{n^i n^j \left[ h_{ij}(x^a_{emit})-h_{ij}(x^a_{rec})\right]}{1-k^in_i},
\end{equation}
where $n^i$ is the laser propagation direction, $k^i$ is the GW propagation 
direction, and the spacetime metric $h_{ij}$ is in transverse-traceless gauge
evaluated at the points of emission and reception at the atom. In our case we
will assume that $n^i$ is in the $\pm\hat x$ direction, and define
$\cos(\theta)=k_x$.  Then, assuming the instrument is optimally oriented
for a linearly polarized GW signal $h(t)$, with $t$ being evaluated at a point $x_0$ midway between the two atom clouds, the relevant metric component is $h_{xx}=\sin^2(\theta) h(t-\cos(\theta)(x-x_0)/c).$

The GW terms, $Y_{\pm}$, in (\ref{eq:PhiL}) and (\ref{eq:PhiR}) are related to the GW Doppler shift in (\ref{eq:Dop}) by a time derivative. Taking into account 
the spacetime emission and reception points, we get
$\tilde{Y}_{\pm}=\tilde{y}_{\pm}/(i\omega)=-\frac{1}{2}\tilde{h}D\sin^2(\theta)e^{-i\omega D/2}\mathrm{sinc}(\omega D(1\mp\cos\theta)/2))$.
For the remainder of our analysis, we will consider a GW source with an optimal sky location ($\theta=\frac{\pi}{2}$), for which 
$\tilde{Y}_{\pm}\approx-\tilde h\omega^{-1}e^{-i\omega D/2}\sin{(\omega D/2)}$.

The output of the GW detector in Figure \ref{fig:AIGW} is obtained by differencing the response of the two AIs, $
\Gamma(t)=\gamma_R(t)-\gamma_L(t-D)$, where $\gamma_{L/R}$ is the response given in (\ref{eq:gamma}) from the left and right AIs respectively.  The GW signal is derived from the difference in the response $\tilde\gamma$ from the two AIs, and thus depends on the optical phase difference at the corresponding vertices on the left and right AI, 
$\Delta\Phi(t)\equiv\delta\Phi_R(t)-\delta\Phi_L(t-D)$. Making substitutions from (\ref{eq:PhiL}) and (\ref{eq:PhiR}) and converting to the Fourier domain, the result is
\begin{equation}
\label{eq:DPhiRot}
\Delta\tilde{\Phi}=2i\sin (\omega D) e^{-i\omega D}\left[k\delta\tilde{x}_2-\delta\tilde{\phi}_2+\frac{ikc}{2\omega}\tilde{h}\right].
\end{equation}
Note that in (\ref{eq:DPhiRot}) the contributions from the left laser's intrinsic frequency noise and from the Doppler shifts induced by the left laser platform's motion are cancelled out but the corresponding terms from the right laser and laser platform remain.
The output of the complete two-cloud GW detector in Figure \ref{fig:AIGW} can be computed by combining (\ref{eq:DPhiRot}) and (\ref{eq:gammaf}):
\begin{equation}
\label{eq:Gammaf}
\tilde{\Gamma}=\beta\left[k\delta\tilde{x}_2-\delta\tilde{\phi}_2+\frac{ikc}{2\omega}\tilde{h}\right]+\Delta\tilde{\Phi}_{m},
\end{equation}
where $\beta\equiv -8i\sin^2(\omega T/2)\sin (\omega D)e^{-i\omega D}$ and $\Delta\tilde\Phi_m$ represents the combined measurement noise in the two AIs.

For comparison, consider the two-way light interferometer signal, which is formed by differencing the two one-way optical phase measurements given by (\ref{eq:PhiL_L}):
\begin{equation}
\label{eq:Gamma_L}
\tilde{\Gamma}^{(l)}=\beta^{(l)}\left[-\frac{ik}{\sin{(\omega D)}}\delta\tilde{x}_{12}-\delta\tilde{\phi}_2+\frac{ikc}{2\omega}\tilde{h}\right]+\Delta\tilde{\Phi}^{(l)}_{m},
\end{equation}
where $\beta^{(l)}\equiv 2i\sin{\omega D}e^{-i\omega D}$, $\delta\tilde{x}_{12}\equiv\delta\tilde{x}_1-\cos(\omega D)\delta\tilde{x}_2$, and $\Delta\tilde{\Phi}^{(l)}_{m}$ is the combined phase measurement noise, for example due to photon shot noise, in the single light interferometer arm.

The strain sensitivity of a GW detector can be computed by comparing the relative sizes of the GW strain and the noise sources in the detector output. For a general GW detector with frequency response $\tilde{\Gamma}$ and noise sources $\tilde{\theta}_\alpha$, the sensitivity can be computed as
\begin{equation}
\label{eq:GenNoise}
S_{h}=\left|\frac{\partial \tilde{\Gamma}}{\partial \tilde{h}}\right|^{-2}\sum_\alpha \left|\frac{\partial \tilde{\Gamma}}{\partial \tilde{\theta}_\alpha}\right|^2 S_{\theta_\alpha},
\end{equation}
where $S_h$ is the power spectral density of the GW strain equivalent to the detector noise and $S_{\theta_\alpha}$ are the power spectral densities of the noise sources.

This formalism can be applied to the GW detector in Figure \ref{fig:AIGW} using the result in (\ref{eq:Gammaf}), yielding contributions to the sensitivity from intrinsic phase noise, platform position noise, and measurement noise. Alternatively, the sensitivity can be expressed in terms of the intrinsic fractional frequency noise, $S_{\delta\nu}\equiv \omega^2/(2\pi\nu)^{2}\,S_{\delta\phi_2}$, the platform acceleration noise, $S_{a_2}\equiv\omega^4 S_{x_2}$, and measurement noise, which we assume is dominated by the combined atomic shot noise, $S_{shot}=2/\eta\,\mbox{Hz}^{-1}$, $\eta$ being the number of detected atoms:
\begin{equation}
\label{eq:Sai}
S_{h}=4S_{\delta\nu}+\frac{4}{\omega^2 c^2}S_{a_2}+\frac{1}{8(kc D)^2 \sin^4(\omega T/2)\eta}\frac{1}{\mbox{Hz}},
\end{equation}
where we've taken the limit $\omega D\ll 1$, consistent with typical AI GW instrument concepts. The literature on proposed AI GW instruments typically only considers the last term in ($\ref{eq:Sai}$)\cite{Hohensee11}. 

The same procedure can be applied to the single-arm light interferometer described by (\ref{eq:Gamma_L}) with the combined measurement noise being the photon shot noise in the two interference measurements, $S^{(l)}_{shot}=2\hbar\nu/P_{rec}\,\mbox{Hz}^{-1}$, where $P_{rec}$ is the light power received from the far light source,
\begin{equation}
\label{eq:Sai_L}
S_{h}^{(l)}=4S_{\delta\nu}+\frac{4}{\omega^4c^2D^2}S_{a_{12}}+\frac{\hbar}{\pi kcD^2 P_{rec}}.
\end{equation}
It is clear from the first terms in (\ref{eq:Sai}) and (\ref{eq:Sai_L}) that detection of a characteristic strain $h$ with either a single-arm light interferometer or a single-arm AI requires a light source with $S_{\delta\nu}^{1/2}\approx h/2$.  The highest performing cavity-stabilized laser systems, which are limited by thermal noise in the cavity mirror coatings, have $S_{\delta\nu}^{1/2}\approx10^{-15}\cdot(f/ 1\,\mbox{Hz})^{-1/2}\,\mbox{Hz}^{-1/2}$ \cite{Webster_08,Alnis_08}. This is roughly six orders of magnitude above the typical strength of astrophysical GW sources in the milliHertz band. 
Light interferometers typically address this problem by utilizing multiple arms. For a detector with two equal-length orthogonal arms driven by a common laser source, the optical phase terms in each measurement will have the same sign and magnitude while the GW term will have opposite sign due to the quadrupolar signature of the response. Differencing the signals from the two arms cancels optical phase noise while retaining the GW signal. This cancellation can be extended to arrangements with unequal length arms using the Time Delay Interferometry (TDI) technique\cite{Armstrong_99}. 

Because acceleration noise in orthogonal directions is uncorrelated, multiple-arm interferometer designs do not allow light source acceleration noise to be cancelled. However, there are two differences in sensitivity between the AI-based GW detector and the light interferometer equivalent. The first is that the AI is sensitive to the absolute acceleration noise of one of the light sources whereas the light interferometer is sensitive to the relative acceleration noise between the two light sources. The second is that the light interferometer sensitivity has an additional factor of $(\omega D)^{-2}$, which means that for a short detector baseline with a given light-source acceleration noise, the light interferometer will have a higher sensitivity limit (less sensitive to GW signals) than the equivalent AI detector. 
In the case of the AI, the light-source acceleration noise requirements for detecting astrophysical GW sources are independent of the baseline, but nonetheless stringent. For example, to reach a strain sensitivity of $S_h^{1/2}\approx 10^{-21}$ at a frequency of $\omega=(2\pi)\,1\,\mbox{mHz}$ would require a light source acceleration noise less than $S_{a}^{1/2}\sim 10^{-15}\,\mbox{m}/\mbox{s}^2/\mbox{Hz}^{1/2}$. 
It makes sense that this is comparable to the residual acceleration requirement on the drag-free test masses in LISA and LISA Pathfinder \cite{Bender_98,McNamara_08} since $(\omega D)_{LISA} \sim 1$.

\emph{Discussion.---}
In this analysis we have compared the basic gravitational-wave response and sensitivity properties of possible space-based atom intereferometer instruments with analogous laser-interferometer instruments, focusing on two of the classic noise sources, spacecraft reference motion and laser phase noise. These noise sources constrain traditional gravitational-wave mission design, but have generally been given little attention in the discussion of AI-based concepts.

We summarize our results in terms familiar to the laser-interferometer GW community. Each AI cloud functions as a (nearly) freefalling laser phasemeter.  The AI signal results from electromagnetic phase signals which are identical to analogous spacecraft-local phase measurements in a light interferometer link in their responses to both gravitational waves and laser frequency, but differ in their responses to the light-source motion. The atom interferometer shows common mode motion of the two end light sources rather than relative motion.  

Because of this difference, in the AI instrument, acceleration noise of one laser source can be cancelled, and the effect on GW sensitivity of the other becomes independent of the instrument baseline.  Beginning with a LISA-like concept, the use of AI would allow the constellation to be shortened \emph{without increasing} the residual acceleration requirements of the reference point.  A smaller instrument would potentially be more sensitive to higher-frequency gravitational-wave signals.  

To determine whether this potential could be realized requires the resolution of a large number of technical issues which fall beyond the scope of this analysis. The acceleration noise requirement on the atom clouds, for instance, does increase when the arms are shortened.  Many of these technical issues have been carefully studied in the AI community, but detailed requirements for a space-based gravitational-wave mission have not been carefully worked out. Where they are known, the requirements often exceed the current performance of ground-based experiments. We also note that we expect the GW sensitivity limits due to optical phase noise and light source acceleration noise discussed here to be generally applicable to more complex AIs. This is because the competition between the GW signal, optical phase noise, and acceleration noise occurs in the \emph{optical} phase. A more precise measurement of this phase with a more complex AI (e.g. high-momentum transfer atomic beam-splitters, 5-pulse interferometers, etc.) may improve GW sensitivity relative to atom shot noise, but will not improve the sensitivity relative to optical phase noise or acceleration noise. We expect the ideas presented here to be helpful in designing future GW instruments which make the best use of AI technology. 

\section{Acknowledgements}
During the preparation of this manuscript, we were made aware of a related analysis of ground-based atomic interferometer gravitational wave detectors made by Jan Harms. We thank Dr. Harms for providing a copy of his report\cite{Harms11} and find our calculations to be in agreement with his in the appropriate limits. We also thank Holger M\"uller and Jeffrey Livas for helpful discussions and Bernard Kelly for his careful review of the manuscript.  
This work was partially supported by NASA grants 08-ATFP08-0126 and 11-ATP11-0046.

\bibliography{ThorpeBaker}

\begin{thebibliography}{10}

\bibitem{Schutz_99}
B. Schutz, Class. Quant. Grav. {\bf 16},  A131  (1999).

\bibitem{Weber60}
J. Weber, Physical Review {\bf 117},  306  (1960).

\bibitem{Weber67}
J. Weber, Physical Review Letters {\bf 18},  498  (1967).

\bibitem{Hellings78}
R. Hellings, Phys. Rev. D {\bf 17},  3158  (1978).

\bibitem{Detweiler79}
S. Detweiler, The Astrophysical Journal {\bf 234},  1100  (1979).

\bibitem{Abromovici92}
A. Abromovici and et~al., Science {\bf 256},  325  (1992).

\bibitem{Tinto98b}
M. Tinto and J. Armstrong, Phys. Rev. D {\bf 58},  042002  (1998).

\bibitem{Hobbs10}
G. Hobbs and et~al., Class. Quant. Grav. {\bf 27},  084013  (2010).

\bibitem{Bender_98}
P. Bender, K. Danzmann, and the LISA Study~Team, Technical Report No.~MPQ233,
  Max-Planck-Institut fur Quantenoptik, Garching (unpublished).

\bibitem{Astro2010}
{Committee for a Decadal Survey of Astronomy and Astrophysics; National
  Research Council}, {\em New Worlds, New Horizons in Astronomy and
  Astrophysics} (The National Academies Press, Washington, D.C., 2010).

\bibitem{Chiao04}
R.~Y. {Chiao} and A.~D. {Speliotopoulos}, Journal of Modern Optics {\bf 51},
  861  (2004).

\bibitem{Roura06}
A. Roura {\it et~al.}, Physical Review D {\bf 73},    (2006).

\bibitem{Delva06}
P. Delva, M. Angonin, and P. Tourrenc, Physics Letters A {\bf 357},  249
  (2006).

\bibitem{Dimopoulos08b}
S. Dimopoulos {\it et~al.}, Physical Review D {\bf 78},  122002  (2008).

\bibitem{Hogan2011}
J. Hogan {\it et~al.}, Gen. Rel. Grav. {\bf 43},  1953  (2011).

\bibitem{Kasevich92}
M. Kasevich and S. Chu, Applied Physics B {\bf 54},  321  (1992).

\bibitem{Cheinet08}
P. Cheinet {\it et~al.}, IEEE Transactions on instrumentation and measurement
  {\bf 57},  1141  (2008).

\bibitem{Kasevich91}
M. Kasevich and S. Chu, Physical Review Letters {\bf 67},  181  (1991).

\bibitem{Snadden98}
M.~J. Snadden {\it et~al.}, Phys. Rev. Lett. {\bf 81},  971  (1998).

\bibitem{Estabrook75}
F. Estabrook and H. Wahlquist, Gen. Rel. Grav. {\bf 6},  439  (1975).

\bibitem{Hellings81}
R. Hellings, Phys. Rev. D {\bf 23},  832  (1981).

\bibitem{Hohensee11}
M. Hohensee {\it et~al.}, Gen. Rel. Grav. {\bf 43},  1905  (2011).

\bibitem{Webster_08}
S. Webster {\it et~al.}, Physical Review A {\bf 77},  033847  (2008).

\bibitem{Alnis_08}
J. Alnis {\it et~al.}, Physical Review A {\bf 77},  053809  (2008).

\bibitem{Armstrong_99}
J. Armstrong, F. Estabrook, and M. Tinto, The Astrophysical Journal {\bf 527},
  814  (1999).

\bibitem{McNamara_08}
P. McNamara, S. Vitale, and K. Danzmann, Class. Quant. Grav. {\bf 25},  114034
  (2008).

\bibitem{Harms11}
J. Harms, Technical Report No.~LIGO-T1100626-v1-H, Laser Interferometer
  Gravitational Wave Observatory (unpublished).

\end{thebibliography}

\end{document}